\documentclass[%
 aip,
 jmp,%
 amsmath,amssymb,
 reprint,%
]{revtex4-1}

\usepackage{graphicx}
\usepackage{dcolumn}
\usepackage{bm}

\begin{document}

\preprint{AIP/123-QED}

\title{Nexus between directionality of terahertz waves and structural parameters in groove-patterned InAs}

\author{Jong-Hyuk  \surname{Yim}}
 \affiliation{School of Information and Communications, Gwangju Institute of Science and Technology,
Gwangju 500-712, South Korea}

\author{Kyunggu  \surname{Min}}
\affiliation{School of Information and Communications, Gwangju Institute of Science and Technology,
Gwangju 500-712, South Korea}

\author{Hoonil \surname{Jeong}}
\affiliation{School of Information and Communications, Gwangju Institute of Science and Technology,
Gwangju 500-712, South Korea}

\author{Jin-Dong  \surname{Song}}
\affiliation{ Nano-Photonics Research Center, Korea Institute of Science and Technology, Seoul
136-791, South Korea}

\author{Young-Dahl  \surname{Jho}}
\email{jho@gist.ac.kr}

\affiliation{School of Information and Communications, Gwangju Institute of Science and Technology,
Gwangju 500-712, South Korea}

\date{\today}

\begin{abstract}

We have performed terahertz (THz)-time domain spectroscopy in various geometries, for
characterizing the directivity of THz waves emitted from groove-patterned InAs structures. First,
we have identified two transient transport processes as underlying THz emission mechanisms in InAs
epilayers with different thickness. The carrier drift around surface depletion region was
predominant in thin sample group (10-70 nm) while the electronic diffusion was overriding the
oppositely aligned drifting dipoles in thick sample group (370-900 nm) as revealed via amplitude
change and phase reversal. By a combined use of electron-beam lithography and inductively coupled
plasma etching in 1 $\mu$m-thick InAs epilayers, we have further periodically fabricated either
asymmetric V-groove patterns or symmetric parabolic apertures. The THz amplitude was enhanced,
particularly along \textit{line-of-sight} transmissive direction when the groove patterns act as
microscale reflective mirrors periodically separated by a scale of diffusion length.

\end{abstract}

\maketitle

\section{\label{sec:level1}INTRODUCTION}

Over the past two decades, there has been great progress in the generating and utilizing terahertz
(THz) waves for various applications, including imaging and communications.\cite{ma} Among various
THz generation methods,\cite{ki} semiconductor surface emitters combined with pulsed laser
excitations has been one of the well-established methods,\cite{xc, ri} which are described by
either non-linear process\cite{sl} or surge current model\cite{pi}; photo-carrier accelerations due
to the surface field in wide band-gap semiconductors or  dipole currents by the efficient
separation between electrons and holes in narrow band-gap semiconductors (so-called, Photo-Dember
effect).

Indium-based alloys are very attractive as contact-free THz emitters because of the large diffusion
velocity difference between electrons and holes. Especially, InAs and InSb have the much higher
electron mobilities (up to 30,000 cm$^{2}$/V$\cdot$s and 76,000 cm$^{2}$/V$\cdot$s) than hole
mobilities (up to 450 cm$^{2}$/V$\cdot$s and 850 cm$^{2}$/V$\cdot$s).\cite{Nainani} In the case of
infrared pulse excitation ($\lambda$ $\sim$800 nm, the central wavelength of Ti:sapphire laser
technology) on InAs and InSb, the absorption depth ($\sim$100 nm) is much shorter than the
diffusion scale ($\sim$1 $\mu$m) and the large excess energy ($> $ 1 eV) leads to the even higher
mobilities of photo-excited carriers.\cite{Dekorsy} In this context, InAs has been widely used as
the most intense contactless THz  emitter whereas InSb application has been limited probably due to
the suppressed emission amplitude via the scattering of electrons into the L valley.\cite{pi}

Although THz efficiency of InAs is appropriate for contact-free THz applications, directivity of
THz waves was hardly controllable. In general, semiconductor-based THz time-domain spectroscopy
(THz-TDS) under pulsed laser excitations are implemented based on many guiding components such as
parabolic mirrors and lenses along reflective directions. Coordinating many optical components
along the reflective directions leads to difficulty in controlling the direction of THz waves.
Recently, THz studies based on diffusive semiconductors have been destined toward the easier
availability and controllability of diffusions;  THz microscopy system was instrumented with
excellent spatial resolution using an optical fiber coupled with tilted InAs tips\cite{mi} and THz
amplitude enhancement was achieved by integrating periodic metal strip lines on
In$_{0.53}$Ga$_{0.47}$As layers for lateral diffusion currents.\cite{gkl} Furthermore, increased
THz magnitude was reported in magnetic fields up to 8 T at 170 K,\cite{rm} or in microstructured
large-area photoconducting emitter with external electric field up to 120 kV/cm at room
temperature.\cite{ad} In this paper, we have implemented micro-scale groove patterns acting either
as mirrors in IR range (in the case of V-groove patterns) or as grids for lateral symmetry breaking
(in the case of parabolic apertures), aiming for characterizing the corresponding directivity of
THz waves as being measured in reflective, transmissive and lateral detection geometries.

\section{SAMPLES AND SCHEMES}

\begin{figure}[!t]
\centering
\includegraphics[scale=0.28,trim=0 0 0 0]{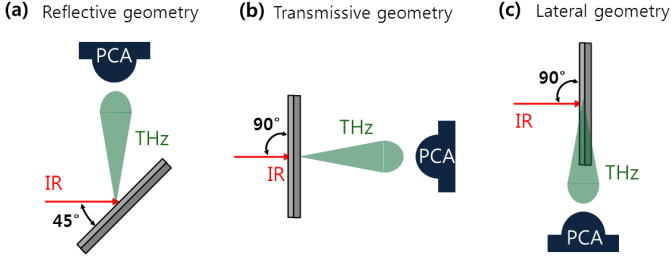}
\caption{Measurement schemes in (a) reflective (b) transmissive, and (c) lateral geometries.}
\end{figure}

The conventional THz-TDS measurements\cite{dg} were performed under the illumination of a
Ti:sapphire-based laser at 300K whose pulse duration was 150 fs pulse at the center wavelength of
800 nm. The samples were coordinated with photo-conductive antennas (PCA) in various detection
geometries for characterizing THz waves along different directions as cartooned in Fig. 1. The IR
laser source was incident on the sample surface at $\theta$=45$\,^{\circ}$ in the reflective
geometry, otherwise at $\theta$=90$\,^{\circ}$ focused onto a spot size of 800 $\mu$m. The pump
beam supplies an excitation fluence of about 0.9 $\mu$J/cm$^2$ which corresponds to the linear
regime in (100)-grown InAs.\cite{Reid} In the case of reflective and transmissive geometries, THz
waves from samples are guided via a pair of off-axis parabolic mirrors into PCA whose sensitivity
was optimized at around 1 THz. On the other hand, in the lateral detection geometry, PCA was placed
very close to the sample (1 mm away from the sample edge) without guiding components.

The samples are classified into two categories; group \textbf{A}) bare InAs epilayers whose
thickness was varied from 10 nm to 900 nm, and group \textbf{B}) 1 $\mu$m-thick samples with
groove-patterns around 1 $\mu$m. The detailed structures and relevant fabrication procedures are
described in section \textbf{III-A} for group \textbf{A} and in in section \textbf{III-B} for group
\textbf{B}, respectively.

\section{RESULTS AND DISCUSSION}

\begin{figure}[!t]
\centering
\includegraphics[scale=0.32,trim=0 0 0 0]{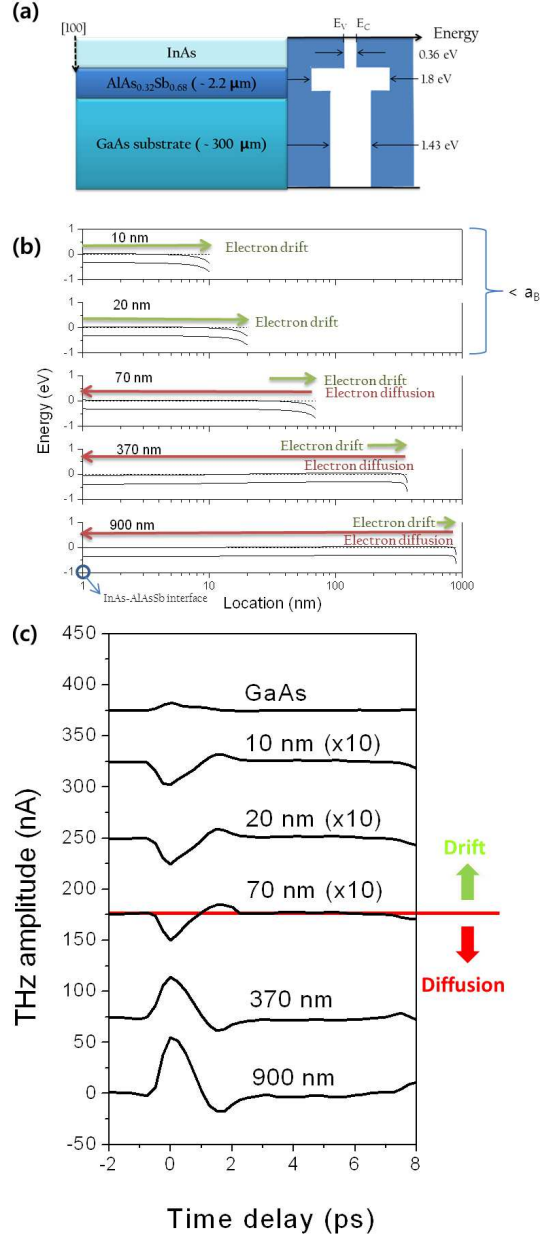}
\caption{(a) InAs epilayers grown on AlAs$_{0.32}$Sb$_{0.68}$ buffer layer and GaAs substrate,
illustrated with the sketchy band gaps. (b) Energy band diagrams of InAs layers as a function of
thickness. Green arrows and red arrows indicate electronic drift or diffusion field directions,
respectively. (c) THz waveforms from InAs layers with different thicknesses, in comparison with
GaAs substrate (from back side illumination of laser). Vertical arrows on the right-hand side
divide THz emission regime, based either on the carrier drift motion (in thin sample group) or on
the carrier diffusion (in thick sample group) with different phase.}
\end{figure}

\subsection{Determination of THz generation mechanism as a function of thickness}

The transient carrier transport mechanisms contribute to the emission of THz waves when surfaces of
semiconductor materials are illuminated by laser pulses; drift currents by photo-carriers around
the surface depletion regions could be biased either towards or against the substrate side,
depending on the doping type, whereas the diffusion (via photo-Dember effect) occurs inherently
toward the substrate side. As discussed earlier, InAs is one of the most efficient contactless THz
emitters, whose generation mechanism has been associated with photo-Dember effect in very thick
samples ($>$ 100 $\mu$m)\cite{xc}. To examine the THz wave generation mechanisms in InAs layer
whose thickness is comparable to or even smaller than the diffusion length scale, we have grown
five unintentionally $n$-type doped InAs epilayers whose thickness was 10, 20, 70, 370, and 900 nm.
To avoid the nonlinear effects, InAs was oriented along (100) direction, grown on GaAs substrate
(thickness $\sim$300 $\mu$m) as shown in Fig. 2(a). To reduce the lattice mismatch and concomitant
strain, AlAs$_{0.32}$Sb$_{0.68}$ layer (thickness $\sim$2.2 $\mu$m) was inserted between InAs and
GaAs substrate. At 300 K, bandgap of materials are roughly illustrated, in parallel with sample
structures on the right-hand side of Fig. 2(a). It is known that the optical rectification effect
is minimized along (100) direction\cite{Liu}, thus, the transport mechanisms are discussed from a
perspective viewpoint of THz wave generation in this and following subsections.

In the energy band diagram of Fig. 2(b) as a function of sample thickness and $n$-type sheet
carrier densities (measured elsewhere)\cite{Jeong}, we found that the electronic drift due to the
tilted potentials at the surface was set to be opposite to the direction of carrier diffusion if
available. In the quantum-confined scale of 10 and 20 nm (thin sample group) which is smaller than
the exciton Bohr radius of 36 nm in a bulk InAs\cite{Fu}, the diffusive features are mostly
suppressed in contrast to the transient electron-hole separation due to the tilted potential shape
around the surface. In 70 nm-thick layer (thin sample group), the direction of transient drift
currents are opposite to that of the partly allowed diffusion currents cancelling the net current
flow.  In thick sample group of 370 nm and 900 nm, the carrier diffusion (mostly dominated by
electrons) can occur throughout the InAs layer, whereas the oppositely aligned drifting electrons
are spatially localized within the surface depletion layer of about 50 nm.

Such thickness-dependent transport features are further confirmed by tracing the amplitudes and
phases of THz waves which are associated with the transient gradient of dipole density, ambient
electric field strength and dipole polarizations. Fig. 2(c) shows the THz-TDS results in reflection
geometry as a function of InAs layer thickness. As a reference, GaAs substrate was illuminated from
backward direction and the corresponding THz waves were displayed on top of Fig. 2(c). The THz
amplitude showed rapid increment after 70 nm. This amplitude variation in samples thicker than the
absorption depth ($\sim$ 100 nm) and exciton Bohr diameter is understood based on the efficient
carrier diffusion and subsequent THz wave radiation. We further note that the opposite direction of
electronic motions between the thin samples (dominated by the carrier drift) and thick sample group
(dominated by the diffusion) led to the opposite phase of THz waves as indicated by the vertical
arrows in Fig. 2(c). Intriguingly, 70 nm-thick sample showed even smaller THz amplitude than
thinner samples, which testifies to the barely initiated diffusions cancelling out the influence
from electron drift along opposite direction.  In this regard, in a thickness range larger than the
370 nm, the underlying THz wave generation mechanisms is based on the electronic diffusion which
will be further engineered for directional control of THz waves in the following section
\textbf{III-B}.

\subsection{Directional THz emission modulated by groove-patterns}

\begin{figure}[!t]
\centering
\includegraphics[scale=0.27,trim=0 0 0 0]{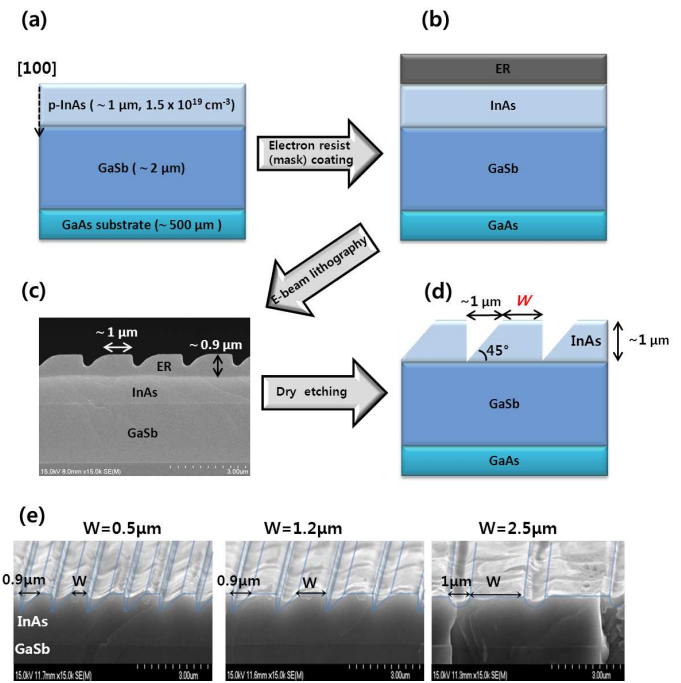}
\caption{ Schematic fabrication processes for forming the groove patterns: (a) the bare InAs
epilayer structure, (b) electron-resist (ER) layer, coated on InAs structure, (c) SEM image after
E-beam writing and developing, and (d) finalized structure by dry etching process. (e) SEM images
of the fabricated structures with different values of $\boldsymbol{W}$ (0.5 $\mu$m and 1.2 $\mu$m
with trapezoidal patterns, and 2.5 $\mu$m with parabolic patterns).}
\end{figure}

InAs epitaxial layers (thickness $\sim$1 $\mu$m) were grown by molecular beam epitaxy along (100)
direction on a GaAs substrate (thickness $\sim$500 $\mu$m) as shown in figure 3(a).  InAs layers
were $p$-type doped with a concentration of $\sim$1.5$\times10^{19}$ cm$^{-3}$. A GaSb layer
(thickness $\sim$2 $\mu$m) was inserted as a buffer layer to reduce the interfacial lattice
mismatch. To fabricate the groove patterns, electron resist (ER) was coated on top of InAs layer
(Fig. 3(b)). The electron beam writing was adapted for developing groove patterns on the ER using
shot-modulation in electron-beam lithography, followed by dry etching process using inductively
coupled plasma etcher with combined use of composite gas of Cl$_{2}$ and N$_{2}$ for forming groove
patterns. Fig. 3(d) shows the proposed structure with a trapezoidal groove angle of 45$\,^{\circ}$
with pattern width ($\boldsymbol{W}$). The pattern width $\boldsymbol{W}$ was varied from 0.5
$\mu$m and 1.2 $\mu$m to 2.5 $\mu$m as shown in the SEM images of Fig. 3(e). It is notable that
$\boldsymbol{W}$ was varied in a comparable scale to the electronic diffusion length in InAs
($\sim$1.3 $\mu$m).\cite{ch} The blue lines in Fig. 3(e) are guide to eyes for intended structures
either with trapezoid-shaped grooves (in the case of $\boldsymbol{W}$=0.5 and 1.2 $\mu$m) or with
parabolic-aperture grooves (in the case of $\boldsymbol{W}$=2.5 $\mu$). During the fabrications,
the edges of groove patterns were slightly deformed, identified as the deviation of SEM images from
the blue lines in Fig. 3(e). The area of groove-patterned region was estimated to be about $5\times
1$ mm$^2$.

Figure 4(a) shows a comparison between THz emission in reflective and in transmissive geometries in
InAs bare samples as a reference. As expected, the signal along the transmissive direction in Fig.
4(a) was mostly suppressed since the diffusive dipoles are mostly aligned along the growth
direction with relatively much weaker lateral components and the subsequent THz radiation patterns
are perpendicular to transient currents. In addition, the THz waves generated within the InAs layer
along the lateral direction are restrictedly coupled with the internal radiation cone along
surface-normal direction due to the large mismatch in refractive index at the surface (the
refractive index of $\sim$3.5 for InAs compared to $\sim$1.0 for free space).\cite{mb} The small
and slow oscillatory features after 10 ps could be ascribed to Fabry-Perot interference between
air-InAs and air-GaAs interfaces.

\begin{figure}[!t]
\centering
\includegraphics[scale=0.42,trim=0 0 0 50]{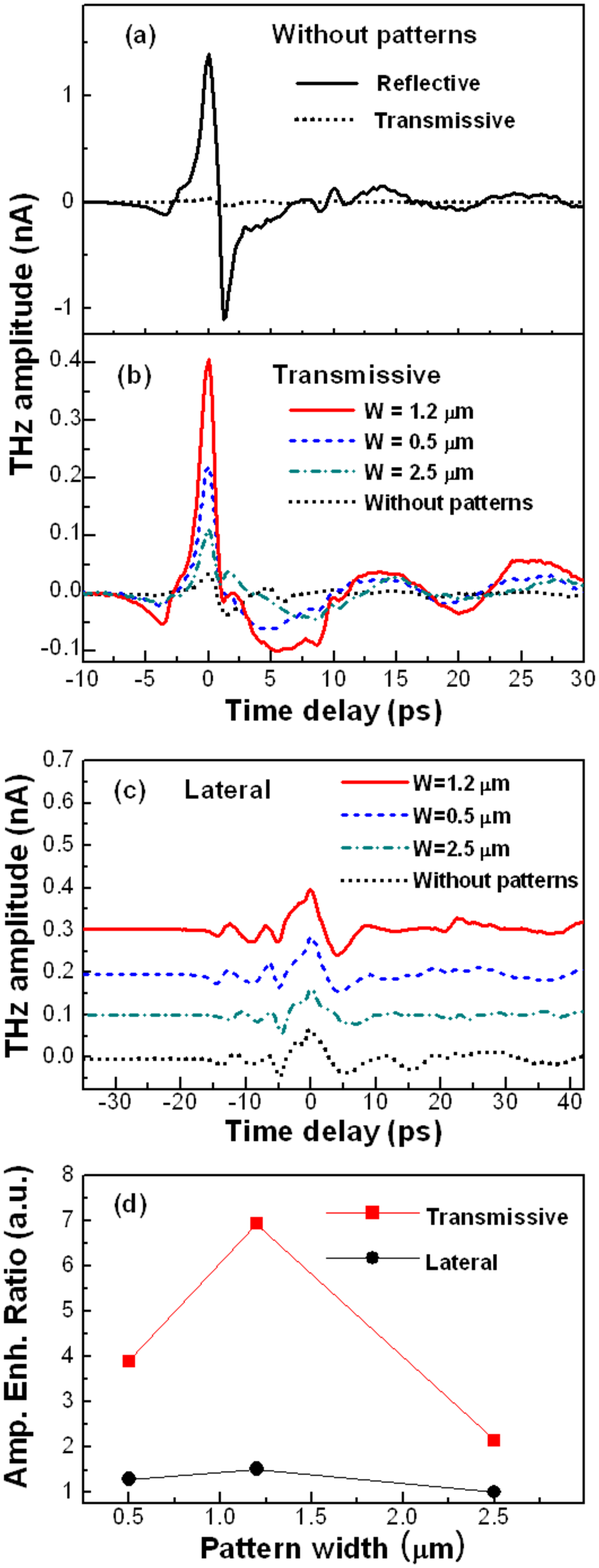}
\caption{Time-domain THz-TDS signals measured from (a) bare InAs layers in the reflective or
transmissive geometries, (b) groove-patterned structures in comparison with the bare InAs layer
(black-dotted line) in the transmissive geometry, and (c) groove-patterned structures in comparison
with the bare InAs layer in the lateral geometry. (d) Amplitude enhancement ratio of THz waves
along transmissive and lateral directions, normalized by amplitudes in bare InAs layer. }
\end{figure}

In the case of groove-patterned structures, in clear contrast to the bare layer, we found that the
THz signals were significantly enhanced along transmissive direction as shown in Fig. 4(b). The
incident beam was normal to the surface and polarized linearly and orthogonally to the striped
patterns. Due to the formation of diffusion-induced transient dipoles along the surface via the
reflected IR pulses toward the lateral direction, the THz radiation patterns are now profiled
toward the line-of-sight direction, manifested by the increased amplitude in Fig. 4(b).
Particularly, a sample with $\boldsymbol{W}$ of 1.2 $\mu$m turned out to be the most efficient
which was in a similar scale to the diffusion length in InAs.\cite{ch} The signals from a sample
with $\boldsymbol{W}$ of 0.5 $\mu$m was less enhanced in that the electrons can easily diffuse from
one side to the other, thus, could be attracted to the rather stationary holes. In the case when
$\boldsymbol{W}$= 2.5 $\mu$m, the patterns could break the lateral symmetry of diffusion but with
limited lateral diffusion efficiency, considering the large laser spot size compared to
$\boldsymbol{W}$. Some further arguments could be raised regarding the results in Fig. 4(b); 1) the
Fabry-Perot interference was persistent in patterned structures, implying that the refractive index
contrast at the interfaces was still persistent. Thus, the radiation outcoupling efficiency between
the inside the materials and the free space seems not influenced by the groove patterns. 2)
photocarriers generated in slope pyramid plane in the trapezoidal patterns could undergo the
diffusion along the (110) direction with the much reduced carrier mobility and with heavier
effective mass compared to those along (100) direction, which leads to the reduced photo-Dember
field and the limited amplitude enhancement. 3) The not-perfectly defined faces of groove patterns
could lead to the scattering loss of the incident IR photons. The pattern scale was much smaller
than the THz wavelength, possibly triggering the Rayleigh scattering. Such arguments could be
further discussed elsewhere based on spectrum-domain studies and angle-dependent amplitude
analysis. Along the lateral direction (Fig. 4(c)), in clear contrast to transmissive direction, we
could not observe the signal changes between patterned structures and the bare samples. The signals
in general in Fig. 4(c) were much weaker than those in other geometries, possibly due to the
slightly misaligned THz guiding optics, the scattering losses at the rough facets, and attenuation
inside the medium. As a comparative purpose, we normalized THz signal amplitudes from the patterned
structures either in the transmissvie or in lateral geometries by those from the bare InAs layer in
Fig. 4(d). THz wave amplitudes were increased significantly depending on the structural parameters
of the periodic groove patterns, only along the line-of-sight trasmission direction. The highest
amplitude enhancement ratio at $\boldsymbol{W}$=1.2 $\mu$m indicates that the optimized adjustment
of $\boldsymbol{W}$ could be further obtained around the diffusion length.

\section{SUMMARY}

For a purpose of directional control of THz waves, 1 $\mu$m-thick InAs epilayers were processed to
have either 45 degree reflectors (with gap width $\boldsymbol{W}$ of 0.5 or 1.2 $\mu$m) or
parabolic apertures with $\boldsymbol{W}$ of 2.5 $\mu$m. The $\boldsymbol{W}$ and the sample
thickness were designed to be similar to the diffusion length in which scale the THz emissions are
predominantly generated by photo-Dember effect. We have performed THz-TDS measurements excited with
a pulsed IR source of Ti:sapphire laser at 300 K. THz-TDS resulted in a dramatic contrast among
different emission directions. In patterned structures, specifically with mirror-like trapezoidal
patterns, THz signals were enhanced along the transmissive direction whereas the lateral THz
emission was not varied by the additional patterning. The enhancement was rather associated with
the efficient lateral dipole formation, optimized when $\boldsymbol{W}$ was similar to the
diffusion length.

\begin{acknowledgments}

This work was supported by the Bio-Imaging Research Center at GIST and by the Basic Science
Research Program through the National Research Foundation of Korea (NRF-2009-0090559). The work at
KIST was supported by the KIST Institutional Program and the Korea-Sweden Research Cooperation
Program.
\end{acknowledgments}


\begin{references}
\bibitem{ma}
M. Tonouchi, Nat. Photonics {\bf 1}, 97 (2007).

\bibitem{ki}
K. Sakai, $Terahertz Optoelectronics$ (Springer, Berlin, 2005).

\bibitem{xc}
X.-C.Zhang, J. T. Darrow, B. B. Hu, D. H. Auston, M. T. Schmidt, P. Tham, and E. S. Yang, Appl. Phys. Lett. {\bf 56}, 2228 (1990).

\bibitem{ri}
R. Ascazubi, I. Wilke, K. Denniston, H. Lu, and W. J. Schaff, Appl. Phys. Lett. {\bf 84}, 4810 (2004).

\bibitem{sl}
S. L. Chuang, S. Schmitt-Rink, B. I. Greene, P. N. Saeta, and A. F. J. Levi, Phys. Rev. Lett. {\bf 68}, 102 (1992).

\bibitem{pi}
P. Gu, M. Tani, S. Kono, K. Sakai, and X.-C. Zhang, J. Appl. Phys. {\bf 91}, 5533 (2002).

\bibitem{Nainani}
A. Nainani, D. Kim, T. Krishnamohan, K. Saraswat, "Hole Mobility and its Enhancement with Strain for Technologically Relevant III-V Semiconductors," Proceedings of SISPAD 2009, pp. 47-50.

\bibitem{Dekorsy}
T. Dekorsy, H. Auer, H. J. Bakker, H. G. Roskos, and H. Kurz, Phys. Rev. B {\bf 53}, 4005 (1996).

\bibitem{mi}
M. Yi, K. Lee, J. Lim, Y. Hong, Y. D. Jho, and J. Ahn, Opt. Express {\bf 18}, 13693 (2010).

\bibitem{gkl}
G. Klatt, B. Surrer, D. Stephan, O. Schubert, M. Fischer, J. Faist, A. Leitenstorfer, R. Huber, and T. Dekorsy, Appl. Phys. Lett. {\bf 98}, 021114 (2004).

\bibitem{rm}
R. McLaughlin, A. Corchia, M.B. Johnston, Q. Chen, C.M. Ciesla, D.D. Arnone, G.A.C. Jones, E.H. Linfield, A.G. Davies, and M. Pepper, Appl. Phys. Lett. {\bf 76}, 2038 (2000).

\bibitem{ad}
A. Dreyhaupt, S. Winnerl, T. Dekorsy, and M. Helm, Appl. Phys. Lett. {\bf 86}, 121114 (2005).

\bibitem{dg}
D. Grischkowsky, S. Keiding, M. van Exter, and C. Fattinger, J. Opt. Soc. Am. B, {\bf 7}, 2006 (1990).

\bibitem{Reid} M. Reid and R. Fedosejevs, Appl. Phys. Lett. {\bf 86}, 011906 (2005).

\bibitem{Liu} K. Liu, J. Xu, T. Yuan and X.-C. Zhang, Phys. Rev. B \textbf{73}, 155330 (2006).

\bibitem{Jeong} H. Jeong, S.H. Shin, S.Y. Kim, J.D. Song, S.B. Choi, D.S. Lee, J. Lee, and Y.D. Jho, Curr. Appl. Phys. {\bf 12}, 668-672 (2012).

\bibitem{Fu}
H. Fu, L.-W. Wang, and A. Zunger, Phys. Rev. B {\bf 59}, 5568 (1999).

\bibitem{ch} C. T. Que, T. Edamura, M. Nakajima, M. Tani, and M. Hangyo, {\bf 48}, 010211 (2009).

\bibitem{mb}
M.B. Johnston, D.M. Whittaker, A. Corchia, A.G. Davies, and E.H. Linfield, Phys. Rev. B, {\bf 65}, 165301 (2002).


\end{references}
\end{document}